\documentclass[DIV=14]{scrartcl}

\title{
Massively parallel read mapping on GPUs\\
with PEANUT}

\author{
Johannes K\"oster\footnote{To whom correspondence should be addressed.}
\and Sven Rahmann}

\date{
Genome Informatics, Institute of Human Genetics\\
University Hospital Essen, University of Duisburg-Essen\\[1ex]
\today}

\usepackage{graphicx}
\usepackage{amsmath,amssymb}
\usepackage{algorithm}
\usepackage{url}
\usepackage[noend]{algpseudocode} 

\newcommand{\qgroups}{\mathcal{G}}
\newcommand{\indexgroup}{I}
\newcommand{\indexgroupstart}{S}
\newcommand{\indexgramstart}{S'}
\newcommand{\indexgramocc}{O}
\newcommand{\qgroupindex}{\mathcal{I}_{T,q}}
\newcommand{\POPCOUNT}{\textsc{Popcount}}
\newcommand{\GROUPRANK}{\textsc{Grouprank}}
\newcommand{\GROUPBIT}{\textsc{Group-And-Bit}}
\newcommand{\INDEXPAIR}{\textsc{Indexpair}}
\newcommand{\kmin}{k_\text{start}}
\newcommand{\kmax}{k_\text{end}}
\newcommand{\given}{\;|\;}

\begin{document}
\maketitle

\begin{abstract}
We present PEANUT (ParallEl AligNment UTility), a highly parallel GPU-based read mapper with several distinguishing features, including a novel q-gram index (called the q-group index) with small memory footprint built on-the-fly over the reads and the possibility to output both the best hits or all hits of a read.
Designing the algorithm particularly for the GPU architecture, we were able to reach maximum core occupancy for several key steps.
Our benchmarks show that PEANUT outperforms other state-of-the-art mappers in terms of speed and sensitivity.
The software is available at \url{http://peanut.readthedocs.org}.
\end{abstract}

\section{Introduction}
\label{section_intro}

A key step in many next generation sequencing (NGS) data analysis projects, e.g.\ for variant calling or gene expression analysis, is mapping the obtained DNA reads to a known reference sequence.
We distinguish between the \emph{read mapping problem}, which is to find the possible origins of each read within the reference, and the \emph{read alignment problem}, which is to provide basepair-level alignments between each read and each originating region.
The read mapping algorithm has to be error tolerant because of mutations and sequencing errors.
The optimal solution to this problem is to calculate semiglobal optimal alignments of each read against the reference.
Since this would incur an infeasible running time (proportional to the product of reference length and total read lengths, e.g. $6\text{~Gbp}\cdot 10^7 \cdot 100\text{~bp} = 6\cdot 10^{18}\text{~bp}^2$ for the human genome and its reverse complement against 10 million Illumina reads), various filtering methods and approximations have been developed.
These can be roughly classified into methods based on backward search using the Burrows-Wheeler transform (BWT), e.g.\ BWA~\cite{li_fast_2009} or Bowtie 2~\cite{langmead_fast_2012}, and methods based on q-gram indexes, e.g.\  RazerS 3~\cite{weese_razers_2012}.

An extension of the read mapping problem is the \emph{split read mapping} problem to also find partial matches of a read, especially of the prefix and suffix of a read.
Split read mapping is useful to detect differential transcript splicing (with the help of reads that span an intron) or structural variations \cite{marschall_clever:_2012}.
One strategy for split read mapping is to partition the read (as done by Tophat \cite{trapnell_tophat:_2009} and others) and map each part with small error tolerance.
If a read mapper is capable of high error tolerance, calculating local alignments can also yield split read mappings (as done by BWA-MEM \cite{li_aligning_2013} and others).

Read mappers can be categorized \cite{weese_razers_2012} into best-mappers that try to find the (or any) best origin of a read (e.g.\ BWA-MEM) and all-mappers that provide a comprehensive enumeration of all possible locations (e.g. RazerS~3 or the ``all'' mode of Bowtie~2) up to a given error threshold.
While all-mappers can be much slower (depending on the number of hits), their strategy is advantageous when a confidence value is required that the reported origin is the true origin of the read.
Further, all-mappers are useful when mapping to homologous sequences like alternative transcripts \cite{roberts_streaming_2013} or meta-genomes, where more than one hit is expected.
An intermediate strategy is to report all hits of the best stratum, i.e., all hits with the same lowest error level (instead of only the first or a random such hit).

Recently, exploiting the parallelization capabilities of graphics processing units (GPUs) for read mapping has become popular and GPU-based BWT-read-mappers appeared, e.g. SOAP3 \cite{liu_soap3:_2012} and SOAP3-dp \cite{luo_soap3-dp:_2013}.
Using a q-gram index on a GPU is not a common choice because of its large size.
Therefore, to the best of our knowledge, q-gram index based mappers so far only use the GPU for calculating the alignments and keep the index on the CPU, e.g.\ NextGenMap \cite{sedlazeck_nextgenmap:_2013} and Saruman \cite{blom_exact_2011}.

Here we present PEANUT (ParallEl AligNment UTility), which is a GPU-based split read mapper using a q-gram index.
PEANUT provides the first feasible implementation of a q-gram index on a GPU, which is achieved by introducing the \emph{q-group index}, a novel q-gram index implementation with a smaller memory footprint, and by indexing subsets of the reads on-the-fly on the GPU.
PEANUT is the first GPU-based all-mapper.
With both a recent and a four years old NVIDIA\textsuperscript{\texttrademark} Geforce GPU, we show that PEANUT outperforms other state of the art best-mappers in terms of speed and sensitivity.
Futher, PEANUT is faster than other all-mappers, while achieving a comparable sensitivity.

This article is structured as follows.
We first lay the foundations for the PEANUT algorithm in Section~\ref{section_foundations} by reviewing requirements for efficient GPU usage and introducing the q-group index data structure.
In Section~\ref{section_algorithm}, we give an overview of the PEANUT algorithm and describe its different steps (filtration, validation, postprocessing) in detail.
Section~\ref{section_results} shows several benchmark results on speed and sensitivity of PEANUT.
A brief discussion concludes the paper.

\section{Foundations}
\label{section_foundations}

PEANUT uses the filtration plus validation approach: It first quickly detects exact matches of a given length~$q$ between each read and each reference (so-called q-gram matches) and only computes alignment scores where such matches are found.
For increasing values of~$q$, this substantially reduces the computational burden.
The advantage of using a $q$-gram index (instead of a compressed suffix array or FM~index) is its simplicity and quick lookup speed.  
The main disadvantage is its high memory usage; we overcome this by indexing a part of the reads on the fly instead of indexing the genome.
We propose a novel implementation of the $q$-gram index functionality (the \emph{q-group-index}) with a smaller memory footprint that is well suited for modern GPU architecture.
We first discuss the GPU architecture and its implications for designing the PEANUT algorithm to maximize parallel GPU usage.
Then, we describe the \emph{q-group index} data structure.

\subsection{Designing for efficient GPU usage}\label{section_gpu_architecture}
We use the terminology of NVIDIA\texttrademark, while the general concepts are also applicable to competitors like AMD\texttrademark.
A GPU is partitioned into Symmetric Multiprocessors (SMs), each of which has its own on-chip memory, cache and processing cores.
By adjusting the \emph{thread block size} it can be controlled how threads are distributed among the SMs.
Each SM executes one thread block until all threads in the block are completed.
Once an SM has completed a thread block, it moves to the next if any blocks are left.
An SM can execute 32~threads in parallel (restricting the thread block size to be a multiple of~32); such a group of threads is called a \emph{warp} or \emph{wavefront}.
At any time, each of these threads has to execute the same instruction in the code, but may do so on different data, a concept which is called \emph{single instruction, multiple threads} (SIMT).
Hence, conditionals with diverging branches should be avoided, since threads taking an if-branch have to wait for threads taking the corresponding else-branch to finish and vice versa.
All SMs may access a slow common global memory (about $\leq 3$~GB on most of today's GPUs) in addition to their fast on-chip cache and memory.
While the size of the fast cache is extremely limited, accesses to the global memory are slow and should be minimized.
However, the memory latency can be reduced by \emph{coalescing} the access, i.e., letting threads in a warp access contiguous memory addresses, such that the same memory transaction can serve many threads.
In addition, an SM can execute a different warp while waiting on a transaction to finish, thereby hiding the latency.
For the latter, threads should minimize their register usage such that the number of warps that can reside on an SM is maximized.
Finally, data transfers from the main system memory to the GPU global memory are comparatively slow.
Hence it is advisable to minimize them as well.

A useful programming pattern that is used extensively in the algorithms presented in this work are \emph{parallel prefix scans}, a special case of which is the computation of a cumulative sum, which at first appears to be a serial process, or filtering an index set.
Parallel prefix scans are used to solve these problems in a data parallel way with a minimum amount of branching, thus nicely fitting the above considerations; see~\cite{blelloch_vector_1990} for a comprehensive introduction.
We use the PyOpenCL implementation of prefix scans~\cite{klockner_pycuda_2012}.


\subsection{The q-group index}

A classical DNA q-gram index of a text~$T$ (usually the reference genome, but for PEANUT a subset of the reads) stores, for each string of length~$q$ over the alphabet, at which positions in~$T$ the q-gram occurs and allows retrieving these positions in constant time per position.
It is commonly implemented via two arrays that we call the \emph{address table}~$A$ and the \emph{position table}~$P$.

Q-grams are encoded as machine words of appropriate size with two subsequent bits encoding one genomic letter (i.e., A = 00, C = 01, G = 10, T = 11).
Unknown nucleotides (usually encoded as N) are converted randomly to A, C, G or T, and larger subsequences of Ns are omitted from the index.
Hence, a q-gram needs $2q$ bits in hardware and is represented (encoded) as a number $g\in \{0,\dots,4^q-1\}$. 
The address table provides for each (encoded) q-gram~$g$ a starting index $A[g]$ that points into the position table such that $P[A[g]], P[A[g]+1], \dots, P[A[g+1]-1]$ are the occurrence positions of $g$.

Deciding about the q-gram length $q$ entails a tradeoff between specificity of the q-grams and the size of the data structure.
Array $A$ needs $4^q$ integers and thus grows exponentially with $q$, while array $P$ needs $|T|$ integers, independently of $q$.
Larger values of $q$ lead to fewer hits per $q$-gram that need to be validated or rejected in later stages.
Further, the choice of~$q$ determines the sensitivity or error-tolerance of the search via the pigeonhole principle (q-gram lemma): only if there are $e < (n+1)/q - 1$ errors, we can guarantee that a q-gram match exists.

\paragraph{Idea.}
The idea of the \emph{q-group index} is to have the same functionality as the q-gram index (i.e., retrieve all positions where a given q-gram occurs in contant time per positions), but with a smaller memory footprint for large~$q$.
This is achieved by introducing additional layers in the data structure.
In the following, we always consider a q-gram as its numeric representation $g\in\{0,\dots,4^q-1\}$.

We divide all $4^q$ q-grams into groups of size~$w$, where $w$ is the GPU word size (typically $w=32$).
The q-gram with the number $g$ is assigned to group number $\lfloor g / w \rfloor$.
Thus the $i$-th group is the set $G_i = \{g \;|\; \lfloor g / w \rfloor = i\}$ of $w$ consecutive q-grams according to their numeric order.
The set of all q-groups is $\qgroups_q := \{ G_0, G_1, \dots, G_{\lceil \frac{4^q}{w} \rceil-1} \}$.
We write $g_{ij}$ for the $j$-th q-gram in $G_i$.

For a given~$q$ and text~$T$, the q-group index is a tuple of arrays
\[ 
\qgroupindex := (\indexgroup, \indexgroupstart, \indexgramstart, \indexgramocc).
\]
Array $\indexgroup$ consists of $|\qgroups_q|$ words with $w$~bits each (overall $4^q$ bits), and bit~$j$ of $\indexgroup[i]$ indicates wheter $g_{ij}$ occurs at all as a substring in the text, i.e.,
\[
\indexgroup[i]_{j} = \begin{cases}
1 &\text{ if } g \text{ is a substring of } T, \\
0 &\text{ otherwise.}
\end{cases}
\]

The array~$\indexgramocc$ corresponds to the position table~$P$ of a regular q-gram array: it is the concatenation of all occurrence positions of each q-gram in sorted numeric q-gram order.
To find where the positions of a particular q-gram~$g$ begin in~$\indexgramocc$, we first determine the group index~$i$ and the q-gram number~$j$ within the group, such that $g=g_{ij}$.
With the bit pattern of $\indexgroup[i]$, we determine whether $q_{ij}$ occurs in $T$.
If not, there is nothing else to do. 
If yes, i.e., $\indexgroup[i]_j=1$, we determine the $j'$ such that bit~$j$ is the $j'$-th one-bit in $\indexgroup[i]$.

\begin{figure}[t]\centering
\includegraphics[width=0.5\textwidth]{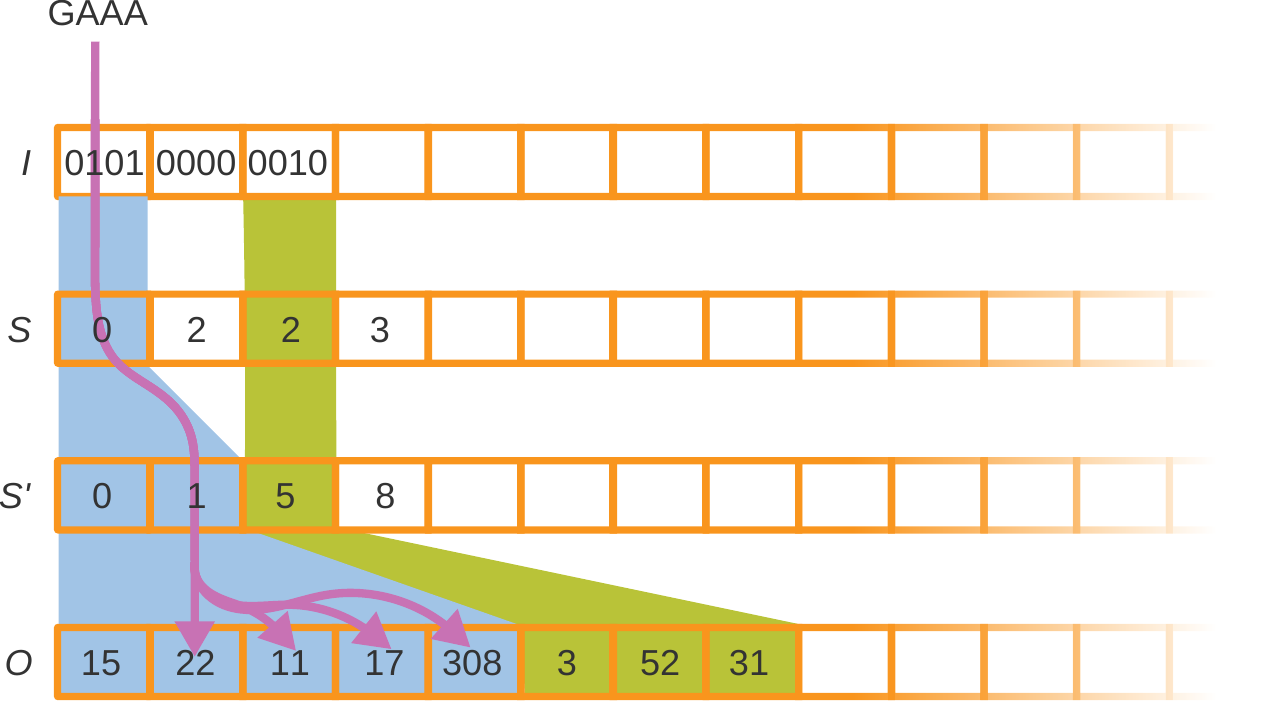}
\caption{The q-group index consists of four arrays $\indexgroup, \indexgroupstart, \indexgramstart, \indexgramocc$.
The purple arrows illustrate how the four layers of the index are traversed to reach the occurrences of the queried q-gram GAAA.}
\label{fig_qgroupindex}
\end{figure}

The address array $\indexgroupstart$ contains, for each group $i$, an index into another address array $\indexgramstart$, such that $\indexgramstart[\indexgroupstart[i] + j']$ is the starting index in $\indexgramocc$ where the positions of $g_{ij}$ can be found; see Figure~\ref{fig_qgroupindex} for an illustration.
All occurrence positions are now listed as
\[ \indexgramocc[\indexgramstart[\indexgroupstart[i] + j']], \dots, \indexgramocc[\indexgramstart[\indexgroupstart[i] + j' + 1]-1].
\]

Similarly to a plain q-gram index, access is in constant time per position:
For a given q-gram~$g$, to determine $(i,j)$ such that $g=g_{ij}$, we simply compute $i=\lfloor g / w \rfloor$ and $j = g-wi = j \bmod w$.
To compute $j'$, i.e., find how many one-bits occur up to bit~$j$ in $I[j]$, we use the population count instruction with a bit mask:
\[ j' = \POPCOUNT(I[i] \,\&\, (2^j - 1)).
\]
The population count $\POPCOUNT(x)$ returns the number of 1-bits in~$x$.

\begin{algorithm}[t]
\begin{algorithmic}[1]
\algblockdefx[Parallel]{Parallel}{EndParallel}[1]{\textbf{for} {#1} \textbf{in parallel}}{\textbf{end parallel}}
\algtext*{EndParallel}

\Require a text $T$, machine word size $w$
\Ensure the q-group index $(\indexgroup, \indexgroupstart, \indexgramstart, \indexgramocc)$

\State initialize $\indexgroup$ with $\lceil 4^q/w \rceil$ zeros
\Parallel{$p \gets 0, \dots, |T|-1$} \label{code_build_step1}
    \State $(i,j) \gets \GROUPBIT(\text{q-gram at position $p$ in $T$})$
	\State $I[i]_j \gets 1$
\EndParallel

\State allocate $\indexgroupstart$ with space for $|\indexgroup|$ integers
\Parallel{$i \gets 0,\dots,|I|-1$} \label{code_build_step2}
	\State $S[i] \gets \POPCOUNT(I[i])$
\EndParallel
\State $S \gets$ cumulative sum of $S$ \label{code_build_step3}

\State initialize $S'$ of length $S[|I|-1]$ with zeros
\Parallel{$p \gets 0, \dots, |T|-1$} \label{code_build_step4}
    \State $(i,j) \gets \GROUPBIT(\text{q-gram at position $p$ in $T$})$
	\State $j' \gets \GROUPRANK(I, i, j)$
	\State increment $S'[S[i] + j']$ by $1$
\EndParallel
\State $S' \gets$ cumulative sum of $S'$ \label{code_build_step5}

\State Allocate $O$ of length $|T|$
\Parallel{$p \gets 0, \dots, |T|-1$} \label{code_build_step6}
    \State $(i,j) \gets \GROUPBIT(\text{q-gram at position $p$ in $T$})$
	\State $j' \gets \GROUPRANK(I, i, j)$
	\State $k \gets$ next free entry in $O$ after $S'[S[i] + j']$
	\State $O[k] \gets p$
\EndParallel
\end{algorithmic}
\caption{Building the q-group index.
For a q-gram~$g$, the function $\GROUPBIT(g)$ computes $(i,j)$ with group $i := \lfloor g/w \rfloor$ and bit $j := g \bmod w$.
The function $\GROUPRANK(I,i,j)$ computes $j':=\text{popcount}(I[i] \,\&\, (2^j-1))$, as explained in the text.}
\label{alg_buildindex}
\end{algorithm}

\paragraph{Construction.}
Algorithm~\ref{alg_buildindex} shows how the index is built.
The outline of of the algorithm is as follows.
First, $\indexgroup$ is created from the q-grams of the text (line~\ref{code_build_step1}).
Then, $\indexgroupstart$ is calculated as the cumulative sum over the population counts of $\indexgroup$ (line~\ref{code_build_step2}).
Next, the number of occurences for each q-gram is calculated (line~\ref{code_build_step4}) and $\indexgramstart$ is created as the cumulative sum over these counts (line~\ref{code_build_step5}).
Finally, the q-gram positions are written into the appropriate intervals of $\indexgramocc$ (line~\ref{code_build_step6}).

Each step is implemented on the GPU with parallel OpenCL\footnote{https://www.khronos.org/opencl} kernels.
The cumulative sums are implemented with parallelized prefix scan operations (see Section~\ref{section_gpu_architecture}).
Importantly, the algorithm needs hardly any branching (hence maximizing concurrency) and makes use of coalescence along the reads in order to minimize memory latency.
All major data structures are kept in GPU memory. 

\paragraph{Size.}
To determine the size of the q-group index, we note that both $\indexgroup$ and $\indexgroupstart$ consist of $\lceil 4^q/w \rceil$ words, $\indexgramstart$ contains an index for each occurring q-gram and hence of $\min\{4^q,|T|\}$ words, and $\indexgramocc$ is a permutation of text positions consisting of $|T|$ words.
Thus the q-group index needs up to $2/w \cdot 4^q + \min\{4^q, |T|\} + |T|$ words, compared to a conventional q-gram index with $4^q + |T|$ words.

If $4^q \lessapprox |T|$, the conventional q-gram index has a small advantage because each q-gram occurs (even multiple times).
In fact, assuming $w=32$, if $4^q = \epsilon |T|$ for some $\epsilon < 1$, the size ratio between q-group index and q-gram index is $(\epsilon/16 + \epsilon + 1)|T| / ((\epsilon+1)|T|) = 1 + \frac{\epsilon}{16(1+\epsilon)}$.
For $\epsilon=1$ (or $|T|=4^q$), this means a small size disadvantage of 3\% for the q-group index.

If on the other hand, $q$~becomes larger for fixed text size (such that q-grams become sparse), the q-group index saves memory, up to a factor of 16.
In this case, write $4^q = K|T|$ for some $K>1$.
The size ratio is $(K/16 + 1 + 1) |T| / ((1+K)|T|) = (2 + K/16)/(1+K)$, which tends to $1/16$ for large $K$.
The break-even point is reached for $K=16/15$.

In practice, we use $q=16$ because a bit-encoded DNA 16-mer just fills a 32-bit word and $q=16$ offers reasonable error tolerance and high specificity.
In this regime with current GPU memory size, we use $|T|=10^8$ (processing 100 million nucleotides at a time), so the size ratio is $K=42.95$, and the q-group index needs only 10\% of the memory of the conventional q-gram index.

\paragraph{Sampling $\indexgroupstart$.}
Without noticable increase in access time, we can reduce the memory usage of the q-group index further by sampling every second (say, even) position of~$S$ and adding another population count and addition instruction to the code for odd positions.


\section{Algorithm}
\label{section_algorithm}

\begin{figure}[t]\centering
\includegraphics[width=0.4\textwidth]{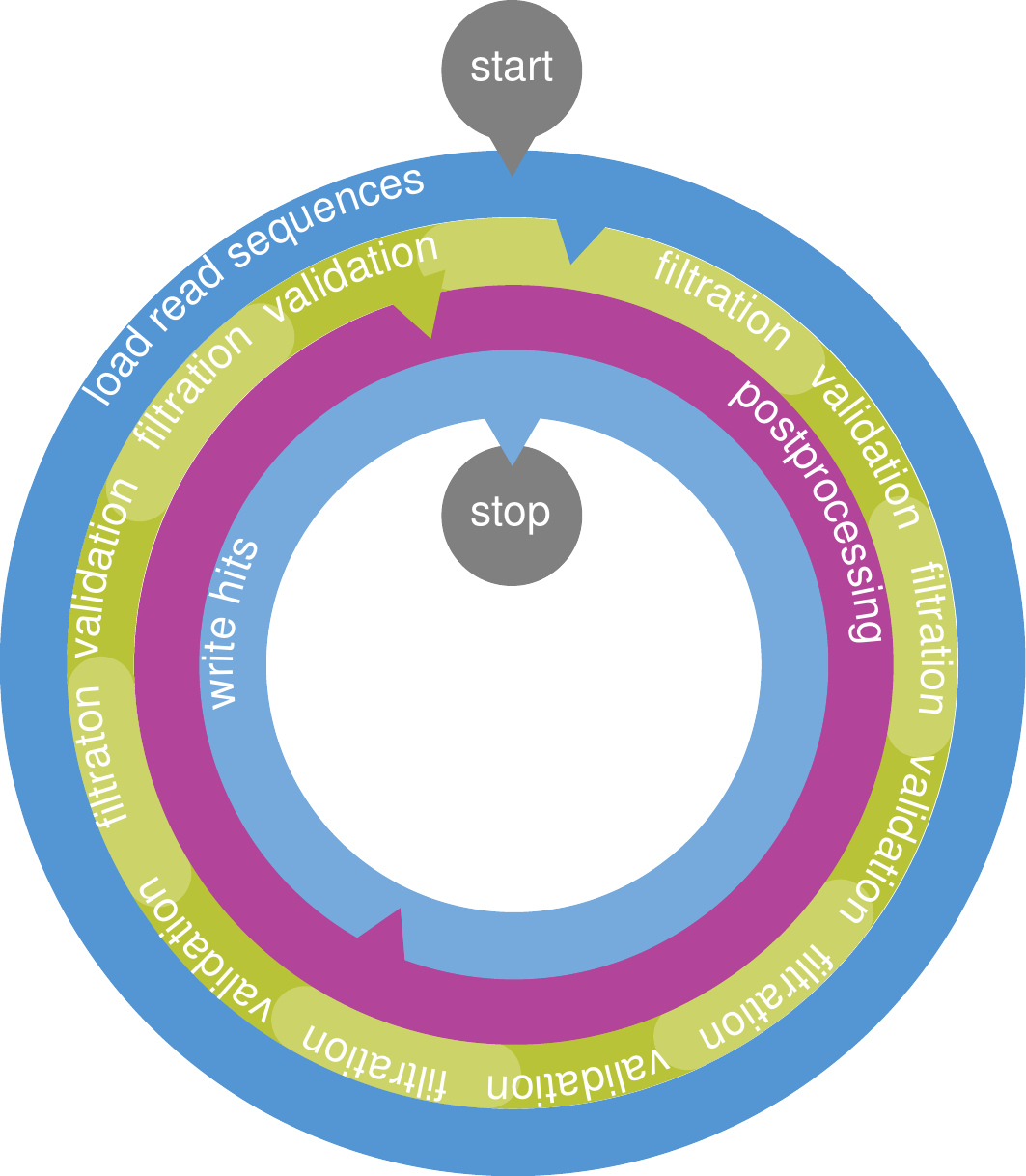}
\caption{The PEANUT algorithm.
Read sequences are buffered and a q-group index is created from them on the fly.
Filtration (detection of q-gram hits) and validation are performed on the GPU until all reference sequences are processed. 
The hits are postprocessed and streamed out in SAM format.
All layers operate independently in parallel and communicate via queues.
Blue layers are I/O bound, green is executed on the GPU, and purple is executed on the CPU.
Arrows between the layers denote a data transfer via a queue.
}
\label{fig_workflow}
\end{figure}

\subsection{Overview}
The PEANUT algorithm consists of the three steps (1)~filtration, (2)~validation and (3)~postprocessing.
The first two steps, filtration and validation are handled on the GPU, while the postprocessing is computed on the CPU.

The steps are conducted on a stream of reads.
Reads are collected until buffers of configurable size are saturated.
Then, any computation is done in parallel for all buffered reads (see Figure~\ref{fig_workflow}).

In the filtration step, potential hits between the reference sequence and the reads are detected using the q-group index.
Next, the potential hits are validated using a variant of Myers' bit-parallel alignment algorithm~\cite{myers_fast_1999, hyyro_bit-vector_2003}.
The validated hits undergo a postprocessing that annotates them with a mapping quality and calculates the actual alignment.
The postprocessed hits are streamed out in SAM format \cite{li_sequence_2009}.
Because of memory constraints on the GPU, all steps are performed per chromosome instead of using the reference as a whole.

\subsection{Filtration}
\label{section_filtration}

The filtration step aims to yield a set of potential hits, e.g. associations between a reference position and a read.
For this, we seek for matching q-grams (i.e. substrings of length $q$) between the reference and the reads.

First, a subset of reads is loaded into memory and its q-group index is created on the GPU on the fly (see Algorithm~\ref{alg_buildindex}).
If the index was built over the reference, either repetetive large data transfers or on-line rebuilding of the index would be necessary for each set of buffered reads and each chromosome.
Hence -- inspired by RazerS~3~\cite{weese_razers_2012} -- we build the q-group index over concatenated reads instead of the reference, so we only build the index once for each set of buffered reads.

We now explain how to use the q-group index to retrieve potential hits.
As stated above, the filtration step will be executed for a buffered set of reads, separately on each reference chromosome.
The reference positions of interest are streamed against the index, and hits are obtained by reference position.

We assume that there is a function $\INDEXPAIR(g)$ that returns, for a q-gram~$g$, an index pair $(\kmin, \kmax)$ such that the occurrence positions in the indexed text are all $\indexgramocc[k]$ with $\kmin \leq k < \kmax$, where $\indexgramocc$ is from the q-group index built with Algorithm~\ref{alg_buildindex}.
Given the q-group index $(\indexgroup, \indexgroupstart, \indexgramstart, \indexgramocc)$, the function $\INDEXPAIR(g)$ is implemented as follows.
Let $(i,j):=\GROUPBIT(g)$ and $j':=\GROUPRANK(I,i,j)$.
Then $\kmin = \indexgramstart[\indexgroupstart[i]+j']$ and $\kmax = \indexgramstart[\indexgroupstart[i]+j'+1]$.

\begin{algorithm}[t]
\begin{algorithmic}[1]
\algblockdefx[Parallel]{Parallel}{EndParallel}[1]{\textbf{for} {#1} \textbf{in parallel}}{\textbf{end parallel}}
\algtext*{EndParallel}

\Require
  reference sequence, 
  set~$P$ of considered reference positions,
  maximum read length $m$, 
  q-group index $(\indexgroup, \indexgroupstart, \indexgramstart, \indexgramocc)$
\Ensure
  array~$H$ of hits as pairs $(d,r)$ of diagonal~$d$ and read id~$r$

\State Initialize array $C$ of length $|P|+1$ with zeros to count hits

\Parallel{$p \in P$} 
\label{code_queryindex_step1}
    \State $(\kmin, \kmax) \gets \INDEXPAIR(\text{q-gram at reference position }p)$
	\State $C[p+1] \gets \kmax - \kmin$
\EndParallel

\State $P \gets \{p \in P \;|\; C[p+1] > 0\}$ 
\label{code_queryindex_step2}

\State $C \gets$ cumulative sum of $C$ 
\label{code_queryindex_step3}

\State Initialize Array $H$ of length $2C[|P|]$ with zeros to store hits

\Parallel{$p \in P$}
\label{code_queryindex_step4}
    \State $(\kmin, \kmax) \gets \INDEXPAIR(\text{q-gram at reference position }p)$
	\For{$k \gets 0 \dots, \kmax-\kmin-1$}
	    \State $p' \gets O[\kmin+k]$
		\State $r \gets \lfloor p' / m \rfloor$ \label{code_queryindex_readid}
		\State $d \gets p - (p' \bmod m)$ \label{code_queryindex_hitpos}
		\State $H[C[p] + k] \gets (d, r)$
	\EndFor
\EndParallel
\end{algorithmic}
\caption{Filtration of reference positions.}
\label{alg_queryindex}
\end{algorithm}

Algorithm~\ref{alg_queryindex} shows how putative hits are generated by querying the q-group index of buffered reads for each q-gram of the reference.
First, the number of hits per reference position are computed in parallel and stored in array $C$ (loop in line~\ref{code_queryindex_step1}).
In the following, only positions with at least one hit are considered (line~\ref{code_queryindex_step2}).
The cumulative sum of the counts generates an interval for each position, that determines where its hits are stored in the output array of the algorithm (line~\ref{code_queryindex_step3}).
Finally the occurences for each reference q-gram are translated into hits that are stored in the corresponding interval of the output array (loop in line~\ref{code_queryindex_step4}).
We translate the position inside the index of concatenated reads into a read number (line \ref{code_queryindex_readid}) and a ``hit diagonal'' that denotes the putative start of the read in the reference (line \ref{code_queryindex_hitpos}, see \cite{rasmussen_efficient_2006}).

Again, each step of Algorithm~\ref{alg_queryindex} is implemented on the GPU with parallel OpenCL kernels.
The filtering of $P$ (line~\ref{code_queryindex_step2}) uses parallelized prefix scan operations (see Section~\ref{section_gpu_architecture}).
All data structures reside in GPU memory; between the steps, at most constant amounts of data have to be transferred between host and GPU (e.g., a single integer).

The set~$P$ of reference positions to investigate and the reference sequence are retrieved from a precomputed HDF5-based reference index\footnote{Hierarchical data format version 5, \url{http://www.hdfgroup.org/HDF5}}.
First, this speeds up access to the reference.
Second, we can pre-compute repetitive regions (defined as exceedingly frequent q-grams) and omit them from the set~$P$.
Such a repeat masking is common in order to avoid uninformative hits when using q-gram index based algorithms.
Finally, $P$~is sorted in numerical order of the q-grams.
This increases the memory coalescence when accessing the q-group index, since subsequent threads will have a higher probability to access the same region in the index and hence the same memory bank in the global GPU memory.

\subsection{Validation}
\label{section_validation}

The validation step takes the potential hits of the filtration step and calculates the edit distance (with unit costs, i.e., the Levenshtein distance) between a read and the reference sequence at its putative mapping position.
If the edit distance is small enough, the hit is considered to be correct and will be postprocessed in the next step.

The edit distance is calculated with Myers' bit-parallel algorithm \cite{myers_fast_1999} that simulates the edit matrix~$E$ between reference sequence and read, which contains one column for each reference base and one row for each read base.
The value $E_{ij}$ is the minimal edit distance between the read prefix of length~$i$ and any substring of the reference that ends at position~$j$.
Interpreted as a graph with a node for each matrix entry, a path in~$E$ from the top to the bottom row denotes a semi-global alignment between read and reference.
Myers' algorithm calculates the edit matrix column-wise.
The state of the current column is stored in bit vectors.
A transition from one column to the next is achieved via a constant amount of bit-parallel operations on the bit vectors.
In iteration~$j$, the minimal distance between the read and any substring of the reference that ends at position~$j$ can be retrieved.
If the accepted error rate is limited, only a part of the edit matrix is needed to calculate the optimal edit distance.
Our implementation of the algorithm follows a banded version that calculates only the relevant diagonal band of the edit matrix \cite{weese_razers_2012}.
The implementation keeps the considered part of each column in a single machine word of size~$w$ (currently 32~bits).
Thereby it provides a complexity of $\mathcal{O}(|r|)$ with $|r|$ being the read length.
While the reduction to the diagonal band restricts the maximum insertion or deletion size in a single alignment, mismatches are not affected.
Hence, the procedure allows to discover partial matches of the read, as needed for split read mapping.
Large indels can be rescued later when calculating the actual alignment if a sufficiently large portion of the read aligns in this step.

Similar to \cite{weese_razers_2012} we use the algorithm to calculate the edit distance of a semi-global alignment in backward direction, thereby obtaining the best starting position of the alignment.
For each hit, the fraction of matches or \emph{percent identity} is obtained from the edit distance as $(|r| - k) / |r|$, where $|r|$ is the read length and $k$ is the edit distance.
Hits with a percent identity less than a given threshold are discarded.
The default for this threshold is 80 percent which provides a decent sensitivity in our benchmarks (see Section~\ref{section_results}).
Decreasing it has moderate impact on performance, since more hits have to be postprocessed and written to disk.

\subsection{Postprocessing}
\label{section_post}
The postprocessing of a read removes duplicate hits (generated by clusters of matching q-grams), sorts the remaining hits by percent identity (our alignment score), pairs mates in case of paired-end alignment, estimates a mapping quality and calculates the actual alignment of each hit.
In contrast to the previous steps, it is done on the CPU.
This allows to postprocess the hits in parallel to filtration and validation.

Intuitively, a particular hit is more likely to be the true origin of a read the fewer hits with the same or with a better score occur.
During postprocessing, hits are sorted into strata of the same percent identity~\cite{marco-sola_gem_2012}.
In paired-end mode, the percent identities of properly paired hits are summed when determining strata.
Upon invocation, PEANUT can be configured to discard hits based on their stratum, e.g. providing only the best stratum or all strata.
In the following, we refer to these as \emph{best-stratum} and \emph{all} modes.

A useful measure for distinguishing true positive hits (i.e. hits referring to the true origin of a read) from false positives is to calculate for each hit the number of hits with a percent identity at least as high.
For read $r$ with $s_{r,p}$ being the best percent identity for an alignment starting at any arbitrary reference position $p \in P$, we denote this as \emph{hit-rank} $R_{r,p}$ and approximate it by
$$R_{r,p} = |\{p' \in P \given s_{r,p'} \geq s_{r,p}\}| \approx |\{p' \in H_r \given s_{r,p'} \geq s_{r,p}\}|$$
with $H_r$ being the validated hits of read~$r$.
The hit-rank can be obtained effectively via sorting $H_r$ by score.
The intuition is that the smaller the hit-rank is (i.e. the higher the percent identity), the more likely the hit will be a true positive.
In Section \ref{section_results} we show that the hit-rank provides a sharp estimate to distinguish between true and false positives.
An important property of the hit-rank is that it is not affected by the applied percent identity threshold since only better hits are considered for its calculation.

PEANUT provides its results in the SAM format \cite{li_sequence_2009}, which expects a mapping quality for each hit.
The mapping quality of read~$r$ at position~$p$ is expected to be the PHRED-scaled error probability, i.e., $-10 \log_{10}$ of the probability that the read does not originate from position~$p$.
To obtain a mapping quality from the hit-rank, we first assume that reads are uniformly sampled from the reference.
Under this null-hypothesis, the probability to obtain by chance a score $X$ at least as extreme as $s_{r,p}$ can then be estimated by
$$\Pr(X \geq s_{r,p}) = \frac{R_{r,p} - 1}{|P|}.$$
Finally, the PHRED-scaled mapping quality is obtained as $\min\{-10 \log_{10} \Pr(X \geq s_{r,p}), 255\}$.


\section{Results}
\label{section_results}

We evaluate PEANUT in terms of its efficiency of GPU resource usage, its sensitivity and its speed.
We also evaluate the ability of the mapping quality measure defined in Section~\ref{section_post} to separate true hits from others.

\begin{figure}\centering
\includegraphics[width=0.4\columnwidth]{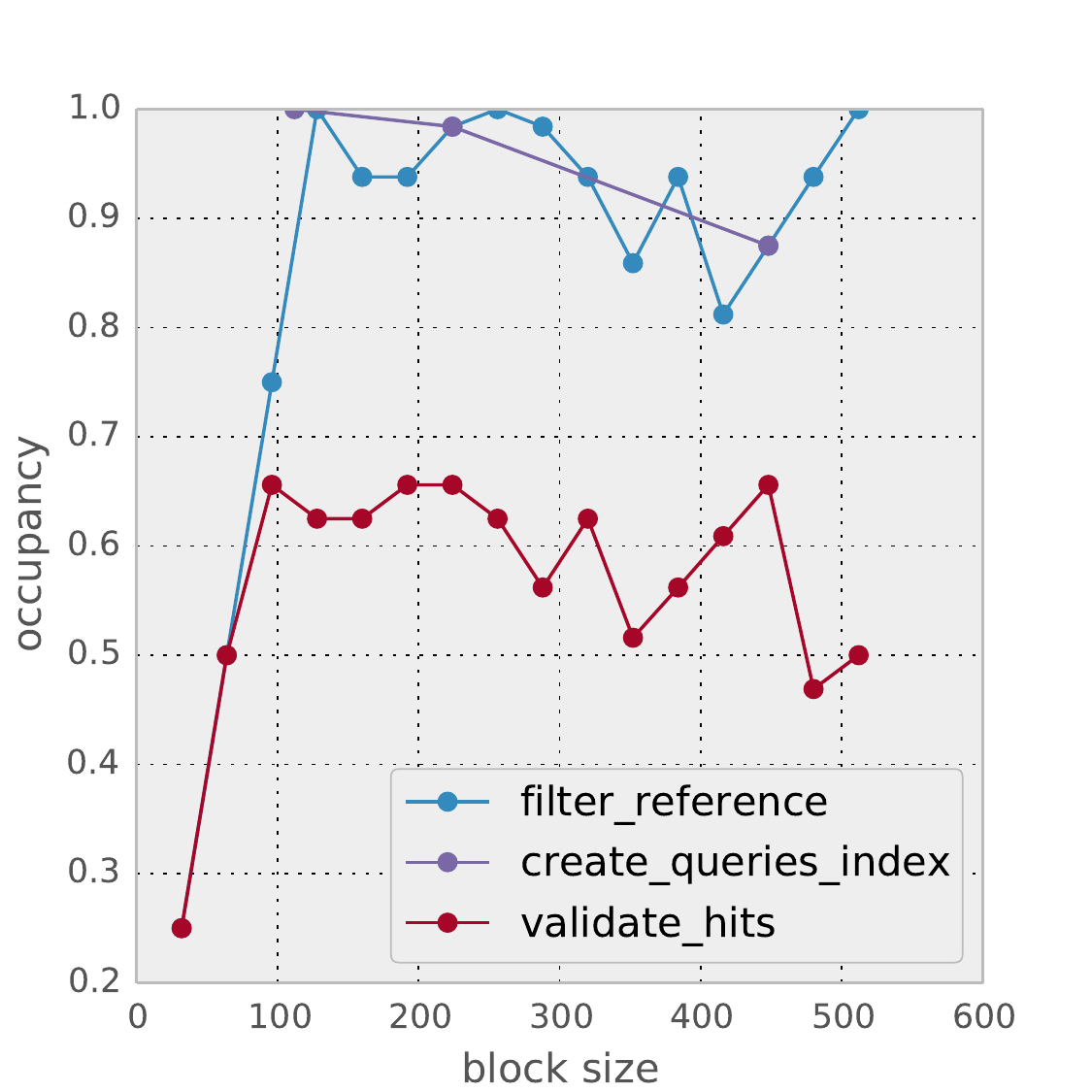}
\caption{The occupancy of GPU cores depending on the thread block size. Shown are representative patterns for OpenCL kernels from the three main steps of the algorithm: building the index (create\_queries\_index), filtration (filter\_reference) and validation (validate\_hits).}
\label{fig_occupancy}
\end{figure}

\paragraph{GPU resource usage.}
In order to maximize utilization of the GPU hardware, idle cores have to be avoided.
The two most important reasons for idle cores are branching and memory latency (see Section \ref{section_gpu_architecture}).
The latter can be hidden if the SM is able to execute a different warp while waiting on the transaction.
The capability to do so can be measured with the \emph{occupancy}, that is the fraction of active warps among the maximum number of warps on an SM.
The more active warps exist on an SM, the higher is the chance that latency can be hidden by executing another warp.
Figure \ref{fig_occupancy} shows the occupancy patterns of the implemented OpenCL kernels, as measured with the NVIDIA\textsuperscript{\texttrademark} CUDA command line profiler depending on the used thread block size (see Section \ref{section_gpu_architecture}).
The thread block size influences the occupancy by limiting the number of potentially active warps and determining the amount of used registers and shared memory on the SM.
Since the latter are limited, a bigger thread block size does not necessarily lead to a higher occupancy.
As can be seen, the occupancy for all steps is high.
For building of the q-group index (create\_queries\_index) and the filtration step (filter\_reference), it even reaches $1.0$ which illustrates the benefit of the q-group index being tailored toward the GPU architecture.

\paragraph{Sensitivity.}
To assess the sensitivity of the algorithm we use the Rabema \cite{holtgrewe_novel_2011} benchmark that allows to compare mapping results based on a formalized framework.
First, 1000 Illumina\footnote{\url{http://www.illumina.com}} reads of length 100 were simulated using Mason \cite{holtgrewe_mason_2010} with the Saccharomyces cerevisiae genome\footnote{as provided by the Rabema data package, \url{http://www.seqan.de/projects/rabema/}, October 2013}, default parameters and error rates.
Second, the simulated reads were mapped to the genome using RazerS~3 \cite{weese_razers_2012} with full sensitivity and different error tolerances.
The mapped reads were used to generate gold standards for Rabema to test against.

We analyze the sensitivity of our algorithm using q-grams of length~16, because it is computationally optimal on the current GPU hardware.
We configure PEANUT to provide all semi-global alignments of a read and leave all other parameters at their default values.
Sensitivity is assessed by the Rabema measure ''Normalized found intervals`` \cite{holtgrewe_novel_2011} and all alignments of a read are considered (all-mode).
With a percent identity threshold of 60 (see Section \ref{section_validation}) our algorithm provides 100\% sensitivity for error rates below 7\%, at least 99.77\% sensitivity for error rates up to 10\% and still 98.96\% sensitivity with an error rate up to an unrealistically high 20\%.
With a stricter threshold of 80, PEANUT still reaches 98.89\% sensitivity for the latter.

In general, we advise to set the percent identity threshold to be slightly more permissive than the expected error rate.
This is because the replacement of N-characters in the reads and the reference with random bases can introduce additional mismatches.
Above rates are far better than the worst case sensitivity that can be expected by applying the pigeonhole principle (i.e., with reads of length~100 and q-grams of length~16, we can expect to find at least one perfectly matching q-gram for all alignments with at most 6~errors), such that using 16-grams appears to be a reasonable default choice in practice.

\paragraph{Running time vs.\ sensitivity.}
We evaluate the run time performance using a dataset of 5~million simulated Illumina HiSeq reads (created with Mason \cite{holtgrewe_mason_2010} at default error rates) on the ENSEMBL human reference genome\footnote{\url{ftp://ftp.ensembl.org/pub/release-74/fasta/homo_sapiens/dna/}} version~37.
Run time performance was measured three times as the overall wall clock time and the best result was selected (best of~3). 
The sensitivity of the alignments (i.e. the percentage of rediscovered alignments from the simulation) was assessed by using the Rabema measure ''Normalized found intervals`` calculated on the simulated data.
Rabema has been run in ''oracle mode`` with the ''any-best`` benchmark type \cite{holtgrewe_novel_2011}, using the original position information from the read simulation.
This means, that the benchmark counts each alignment that is equally good as the one from the read simulation as correct.
We consider this benchmark to be fair, since the best mappers BWA-MEM and Bowtie~2 per default only provide a single alignment even if there are multiple ones of the same quality.

The benchmark was conducted on an Intel Core i7-3770\textsuperscript{\texttrademark} system (4 cores with hyperthreading, 3.4GHz, 16 GB RAM), an NVIDIA Geforce\textsuperscript{\texttrademark} 780 GPU (12 SMs, 3GB RAM) and a 7200 rpm hard disk.
We evaluated two modes of PEANUT.
First, PEANUT was configured to find the best stratum of semi-global alignments (best-stratum mode) for each read.
Second, PEANUT was configured to find all semi-global alignments (all mode) for each read.
Further, we benchmarked the newest generation of BWA (BWA-MEM,~\cite{li_aligning_2013}), Bowtie~2 \cite{langmead_fast_2012} and RazerS~3 \cite{weese_razers_2012}.
(At the time of writing, we were not able to obtain a working binary or source distribution of the GPU based read mappers SOAP3 \cite{liu_soap3:_2012} and SOAP3-dp \cite{luo_soap3-dp:_2013} for our setup.
Further, NextGenMap \cite{sedlazeck_nextgenmap:_2013} was not able to detect our OpenCL setup, and therefore refused to run.)
While BWA-MEM and Bowtie~2 provide only the best hit, RazerS~3 per default reports all possible alignments.
All read mappers were configured to output alignments in SAM format \cite{li_sequence_2009} directly to the hard disk.

\begin{table}[t]\centering
\caption{Performance and sensitivity of PEANUT and other state of the art read mappers on the human reference genome with (a) 5~million simulated Illumina reads of length 100, (b) 5~million real Illumina HiSeq 2000 reads from the human exome and (c) 10~million paired-end real Illumina HiSeq 2000 reads from the human exome.
For the simulated dataset, sensitivity was measured with Rabema \cite{holtgrewe_novel_2011}.
Running time is given for the quickest run of three repetitions.
}\label{table_performance}

\begin{tabular}{rccrr}
dataset & mapper & type & time [min:sec] & sens.\ [\%]\\
\hline\hline
a & PEANUT & best-stratum & \textbf{1:55} & \textbf{98.62}\\
a & BWA-MEM & best & 3:16 & 96.99\\
a & Bowtie 2 & best & 5:21 & 96.85\\
\hline
a & PEANUT & all & \textbf{18:29} & 98.74\\
a & RazerS 3 & all & 199:55 & \textbf{98.83}\\
\hline\hline
b & PEANUT & best-stratum & \textbf{1:33} & N/A\\
b & BWA-MEM & best & 1:58 & N/A\\
b & Bowtie 2 & best & 3:12 & N/A\\
\hline
b & PEANUT & all & \textbf{10:52} & N/A\\
b & RazerS 3 & all & 89:38 & N/A\\
\hline\hline
c & PEANUT & best-stratum & \textbf{3:08} & N/A\\
c & BWA-MEM & best & 4:44 & N/A\\
c & Bowtie 2 & best & 8:18 & N/A\\
\hline
c & PEANUT & all & \textbf{21:54} & N/A\\
c & RazerS 3 & all & 150:59 & N/A\\
\hline\hline
\end{tabular}
\end{table}

Table~\ref{table_performance} compares PEANUT with its competitors.
First, PEANUT in best-stratum mode outperforms the best-mappers BWA-MEM and Bowtie~2 in terms of speed and sensitivity.
Second, PEANUT in all-mode is 10 times faster than the all-mapper RazerS~3, while maintaining a comparable sensitivity.
While Bowtie~2 provides an all-mode, too, it did not terminate in competetive time due to extensive memory requirements exceeding the capabilities of our test system.

In addition, we measured the run times using 10~million paired-end reads of of the publicly available exome sequencing dataset ERR281333\footnote{\url{http://www.ebi.ac.uk/ena/data/view/PRJEB3979}} \cite{martin_exome_2013} obtained from an uveal melanoma sample sequenced with Illumina HiSeq 2000.
When interpreting the 5~million forward direction reads of above dataset as single-end reads, the run time results resemble those of the simulation benchmark, but all four mappers perform slightly faster.
This is likely caused by the greater locality of data structure accesses caused by the exome-sequenced reads, which should be less distributed across the genome.
Nevertheless, PEANUT is again faster than its competitors in both modes.
Finally, we investigated the paired-end performance of all mappers using above dataset.
Again, PEANUT proves to be faster than the competitors.

\begin{table}[t]\centering
\caption{
Performance of PEANUT on secondary test system with a four years old Geforce~580 GPU vs.\ a more recent Geforce~780.
Times are in min:sec.
See also Table~\ref{table_performance}.
}\label{table_performance_secondary}

\begin{tabular}{rcrrr}
dataset & type & time (580) & time (780) & difference\\
\hline\hline
a & best-stratum & 2:11 & 1:55 & 13.9\% \\
a & all & 20:56 & 18:29 & 13.3\% \\
\hline
b & best-stratum & 1:45 & 1:33 & 12.9\% \\
b & all & 11:49 & 10:52 & 8.7\% \\
\hline
c & best-stratum & 3:28 & 3:08 & 10.6\% \\
c & all & 23:55 & 21:54 & 9.2 \% \\
\hline\hline
\end{tabular}
\end{table}

While PEANUT requires at least 2.5 GB of GPU memory for filtration and validation, it is not restricted to running on high-end GPU models like the Geforce 780 used above:
Table~\ref{table_performance_secondary} shows that a speedup can be maintained when benchmarking on a different test system with an Intel Core i7-2600 (3.4GHz, 16GB RAM) and a four years old NVIDIA\textsuperscript{\texttrademark} Geforce 580 GPU.
We see that the older system is 9\% to 14\% slower, depending on the dataset and still faster than the best competitor in Table~\ref{table_performance}.

\begin{figure}[t]\centering
\includegraphics[width=0.4\columnwidth]{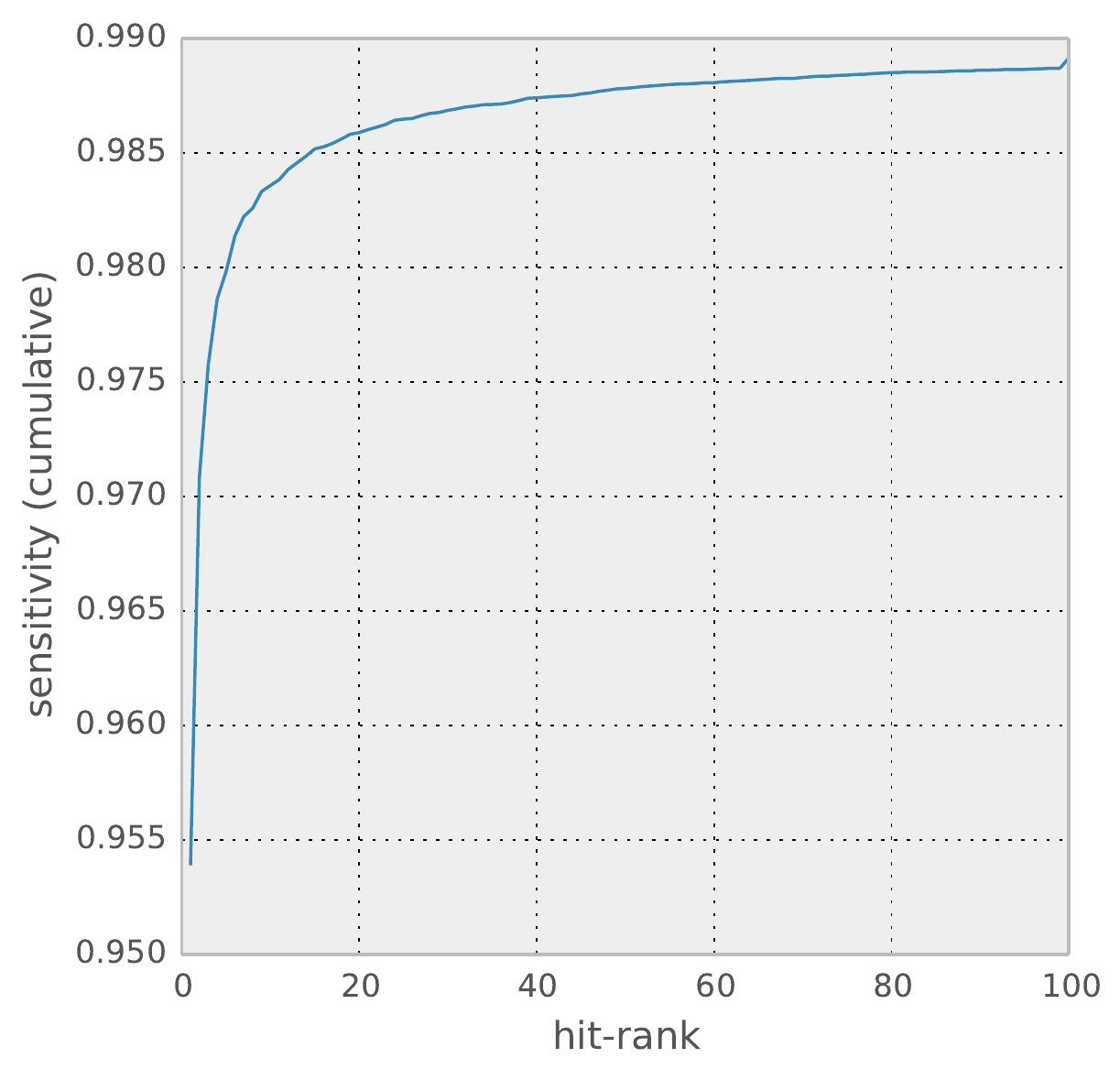}
\caption{Hit-rank vs. sensitivity for 100000 simulated reads from the human reference genome. For each hit-rank $k$, the y-axis depicts the sum of the sensitivities up to~$k$.}\label{fig_hit_rank}
\end{figure}

\paragraph{Hit-rank vs. sensitivity.}
Finally we evaluate the ability of the hit-rank (see Section \ref{section_post}) to distinguish between true and false positives.
Here, true positives are the correctly identified true sampling positions of the reads, whereas false positives are reported mapping locations that may have the same alignment score but are not the true origins of a read.
For this purpose the sensitivity was calculated for each hit-rank in the first 100000 reads of the simulated human dataset from above.
Figure \ref{fig_hit_rank} shows the accumulated sensitivity vs. the considered hit-ranks.
As can be seeen, the hit-rank (and hence the mapping qualities of PEANUT) provides a sharp distinction between true and false positives and small ranks account for most of the sensitivity.

\section{Discussion}

We presented PEANUT, a read mapper exploiting the capabilities of modern GPUs using OpenCL.
PEANUT is based on the q-group index, a novel variant of a q-gram index that has a particularly small memory footprint and is therefore tailored for modern GPUs.
To the best of our knowledge, the presented datastructure and algorithm is the first implementation of the filtration-validation approach that works completely on the GPU.
The benefit of this is illustrated by PEANUT showing supreme speed over state of the art read mappers like BWA, Bowtie~2 and RazerS~3.
This holds for both the high-end GPU NVIDIA\textsuperscript{\texttrademark} Geforce 780 and a four years old NVIDIA\textsuperscript{\texttrademark} Geforce 580.
Thereby, PEANUT maintains a high sensitivity that is comparable to RazerS~3 and higher than BWA and Bowtie~2.

PEANUT is distributed under the MIT license as an open source Python\footnote{\url{http://www.python.org}} software package.
Filtration and validation where implemented in OpenCL, using the PyOpenCL package \cite{klockner_pycuda_2012} and postprocessing was implemented in Cython \cite{behnel_cython:_2011}.
Documentation and installation instructions are available at \url{http://peanut.readthedocs.org}, where we further provide a Snakemake \cite{koster_snakemake_2012} workflow of all analyses conducted in this work.

\paragraph{Acknowledgments.}
Part of this work was funded by the German Research Foundation (DFG), Collaborative Research Center (Sonderforschungsbereich,~SFB) 876 ``Providing Information by Resource-Constrained Data Analysis'' within project C1, see \url{http://sfb876.tu-dortmund.de}.
SR acknowledges support from the Mercator Foundation within project MERCUR Pe-2013-0012 (UAMR Professorship ``Computational Biology'').

\bibliography{peanut}

\begin{thebibliography}{10}

\bibitem{behnel_cython:_2011}
Stefan Behnel, Robert Bradshaw, Craig Citro, Lisandro Dalcin, Dag~Sverre
  Seljebotn, and Kurt Smith.
\newblock Cython: The best of both worlds.
\newblock {\em Computing in Science and Engg.}, 13(2):31–39, March 2011.

\bibitem{blelloch_vector_1990}
Guy~E. Blelloch.
\newblock {\em Vector Models for Data-parallel Computing}.
\newblock {MIT} Press, Cambridge, {MA}, {USA}, 1990.

\bibitem{blom_exact_2011}
Jochen Blom, Tobias Jakobi, Daniel Doppmeier, Sebastian Jaenicke, J\"orn
  Kalinowski, Jens Stoye, and Alexander Goesmann.
\newblock Exact and complete short-read alignment to microbial genomes using
  graphics processing unit programming.
\newblock {\em Bioinformatics}, 27(10):1351--1358, May 2011.

\bibitem{holtgrewe_mason_2010}
Manuel Holtgrewe.
\newblock Mason -- a read simulator for second generation sequencing data,
  October 2010.
\newblock Technical Report {TR-B-10-06}, Institut f\"ur Mathematik und
  Informatik, Freie Universit\"at Berlin.

\bibitem{holtgrewe_novel_2011}
Manuel Holtgrewe, Anne-Katrin Emde, David Weese, and Knut Reinert.
\newblock A novel and well-defined benchmarking method for second generation
  read mapping.
\newblock {\em {BMC} Bioinformatics}, 12(1):210, May 2011.

\bibitem{hyyro_bit-vector_2003}
Heikki Hyyr\"o.
\newblock A bit-vector algorithm for computing levenshtein and damerau edit
  distances.
\newblock {\em Nordic Journal of Computing}, page 2003, 2003.

\bibitem{klockner_pycuda_2012}
Andreas Kl\"ockner, Nicolas Pinto, Yunsup Lee, Bryan Catanzaro, Paul Ivanov,
  and Ahmed Fasih.
\newblock {PyCUDA} and {PyOpenCL:} a scripting-based approach to {GPU} run-time
  code generation.
\newblock {\em Parallel Computing}, 38(3):157--174, March 2012.

\bibitem{koster_snakemake_2012}
Johannes K\"oster and Sven Rahmann.
\newblock Snakemake -- a scalable bioinformatics workflow engine.
\newblock {\em Bioinformatics}, 28(19):2520--2522, October 2012.

\bibitem{langmead_fast_2012}
Ben Langmead and Steven~L. Salzberg.
\newblock Fast gapped-read alignment with bowtie 2.
\newblock {\em Nature Methods}, 9(4):357--359, April 2012.

\bibitem{li_aligning_2013}
Heng Li.
\newblock Aligning sequence reads, clone sequences and assembly contigs with
  {BWA-MEM}.
\newblock {arXiv} e-print 1303.3997, March 2013.

\bibitem{li_fast_2009}
Heng Li and Richard Durbin.
\newblock Fast and accurate short read alignment with {Burrows-Wheeler}
  transform.
\newblock {\em Bioinformatics}, 25(14):1754--1760, July 2009.

\bibitem{li_sequence_2009}
Heng Li, Bob Handsaker, Alec Wysoker, Tim Fennell, Jue Ruan, Nils Homer, Gabor
  Marth, Goncalo Abecasis, and Richard Durbin.
\newblock The sequence {Alignment/Map} format and {SAMtools}.
\newblock {\em Bioinformatics}, 25(16):2078--2079, August 2009.

\bibitem{liu_soap3:_2012}
Chi-Man Liu, Thomas Wong, Edward Wu, Ruibang Luo, Siu-Ming Yiu, Yingrui Li,
  Bingqiang Wang, Chang Yu, Xiaowen Chu, Kaiyong Zhao, Ruiqiang Li, and Tak-Wah
  Lam.
\newblock {SOAP3:} ultra-fast {GPU-based} parallel alignment tool for short
  reads.
\newblock {\em Bioinformatics}, 28(6):878--879, March 2012.

\bibitem{luo_soap3-dp:_2013}
Ruibang Luo, Thomas Wong, Jianqiao Zhu, Chi-Man Liu, Xiaoqian Zhu, Edward Wu,
  Lap-Kei Lee, Haoxiang Lin, Wenjuan Zhu, David~W. Cheung, Hing-Fung Ting,
  Siu-Ming Yiu, Shaoliang Peng, Chang Yu, Yingrui Li, Ruiqiang Li, and Tak-Wah
  Lam.
\newblock {SOAP3-dp:} fast, accurate and sensitive {GPU-Based} short read
  aligner.
\newblock {\em {PLOS} {ONE}}, 8(5):e65632, May 2013.

\bibitem{marco-sola_gem_2012}
Santiago Marco-Sola, Michael Sammeth, Roderic Guigó, and Paolo Ribeca.
\newblock The {GEM} mapper: fast, accurate and versatile alignment by
  filtration.
\newblock {\em Nature Methods}, 9(12):1185--1188, December 2012.

\bibitem{marschall_clever:_2012}
Tobias Marschall, Ivan~G. Costa, Stefan Canzar, Markus Bauer, Gunnar~W. Klau,
  Alexander Schliep, and Alexander Sch\"onhuth.
\newblock {CLEVER:} clique-enumerating variant finder.
\newblock {\em Bioinformatics}, 28(22):2875--2882, November 2012.

\bibitem{martin_exome_2013}
Marcel Martin, Lars Maßh\"ofer, Petra Temming, Sven Rahmann, Claudia Metz,
  Norbert Bornfeld, Johannes van~de Nes, Ludger Klein-Hitpass, Alan~G.
  Hinnebusch, Bernhard Horsthemke, Dietmar~R. Lohmann, and Michael Zeschnigk.
\newblock Exome sequencing identifies recurrent somatic mutations in {EIF1AX}
  and {SF3B1} in uveal melanoma with disomy 3.
\newblock {\em Nature Genetics}, 45(8):933--936, August 2013.

\bibitem{myers_fast_1999}
Gene Myers.
\newblock A fast bit-vector algorithm for approximate string matching based on
  dynamic programming.
\newblock {\em J. {ACM}}, 46(3):395–415, May 1999.

\bibitem{rasmussen_efficient_2006}
Kim~R Rasmussen, Jens Stoye, and Eugene~W Myers.
\newblock Efficient q-gram filters for finding all epsilon-matches over a given
  length.
\newblock {\em Journal of computational biology: a journal of computational
  molecular cell biology}, 13(2):296--308, March 2006.

\bibitem{roberts_streaming_2013}
Adam Roberts and Lior Pachter.
\newblock Streaming fragment assignment for real-time analysis of sequencing
  experiments.
\newblock {\em Nature Methods}, 10(1):71--73, January 2013.

\bibitem{sedlazeck_nextgenmap:_2013}
Fritz~J Sedlazeck, Philipp Rescheneder, and Arndt von Haeseler.
\newblock {NextGenMap:} fast and accurate read mapping in highly polymorphic
  genomes.
\newblock {\em Bioinformatics (Oxford, England)}, 29(21):2790--2791, November
  2013.

\bibitem{trapnell_tophat:_2009}
Cole Trapnell, Lior Pachter, and Steven~L. Salzberg.
\newblock {TopHat:} discovering splice junctions with {RNA-Seq}.
\newblock {\em Bioinformatics}, 25(9):1105--1111, May 2009.

\bibitem{weese_razers_2012}
David Weese, Manuel Holtgrewe, and Knut Reinert.
\newblock {RazerS} 3: Faster, fully sensitive read mapping.
\newblock {\em Bioinformatics}, 28(20):2592--2599, October 2012.

\end{thebibliography}
\bibliographystyle{plain}
\end{document}